\documentclass[acmsmall, nonacm]{acmart}

\usepackage{graphicx}
\usepackage{amsmath}
\usepackage{url}
\usepackage{subcaption}
\usepackage{multirow}
\usepackage{tikz}
\usepackage{relsize}
\usepackage{adjustbox}
\usepackage{colortbl}
\usepackage{comment}
\usepackage{array}
\usepackage{tabularx}
\usepackage{float} 
\usepackage{longtable} 
\usepackage{titlesec}
\usepackage{makecell}
\usepackage{hhline}
\usepackage{array}
\usepackage{makecell}
\usepackage{cleveref}
\usepackage{comment}
\usepackage{mathtools}
\usepackage{soul}
\usepackage{enumitem}
\usepackage{arydshln} 
\usepackage{diagbox} 
\usepackage{soul}
\usepackage{listings}
\usepackage[utf8]{inputenc}
\usepackage{graphicx}

\makeatletter
 \newcommand{\shorteq}{%
   \settowidth{\@tempdima}{-}
  \resizebox{\@tempdima}{\height}{=}%
}
\makeatother

\newcommand{\hide}[1]{}
\newcolumntype{^}{>{\currentrowstyle}}
\newcommand{\rowstyle}[1]{\gdef\currentrowstyle{#1}%
   #1\ignorespaces
}

\newcommand{\thickhline}{%
     \noalign {\ifnum 0=`}\fi \hrule height 1pt
     \futurelet \reserved@a \@xhline
}

\usepackage{xspace}
\newcommand{\etal}{\emph{et al.}\xspace}
\newcommand{\eg}{\emph{e.g}\xspace}

\definecolor{darkred}{rgb}{0.55, 0.0, 0.0}
\definecolor{ao(english)}{rgb}{0.0, 0.5, 0.0}
\definecolor{bluebell}{rgb}{0.64, 0.64, 0.82}
\definecolor{dollarbill}{rgb}{0.52, 0.73, 0.4}
\definecolor{cardinal}{rgb}{0.77, 0.12, 0.23}

\begin{document}

\title{To BEE or Not to BEE: Estimating more than Entropy with \textbf{B}iased \textbf{E}ntropy \textbf{E}stimators}

\author{Ilaria Pia la Torre}
\email{ilaria.torre.20@ucl.ac.uk}
\affiliation{%
    \institution{University College London}
    \city{London}
    \country{UK}}
\orcid{0009-0006-2733-5283}
\author{David A. Kelly}
\email{david.a.kelly@kcl.ac.uk}
\orcid{0000-0002-5368-6769}
\affiliation{%
    \institution{King's College London}
    \city{London}
    \country{UK}
}
\author{H\'{e}ctor D. Men\'{e}ndez}
\email{hector.menendez@kcl.ac.uk}
\orcid{0000-0002-6314-3725}
\affiliation{%
    \institution{King's College London}
    \city{London}
    \country{UK}
}
\author{David Clark}
\email{david.clark@ucl.ac.uk}
\orcid{0000-0002-7004-934X}
\affiliation{%
    \institution{University College London}
    \city{London}
    \country{UK}
}

\begin{abstract}
Entropy estimation plays a significant role in biology, economics, physics, communication engineering and other disciplines. 
It is increasingly used in software engineering, \eg in software confidentiality, software testing, predictive analysis, machine learning, and software improvement. However accurate estimation is demonstrably expensive in many contexts, including software. Statisticians have consequently developed biased estimators that aim to accurately estimate entropy on the basis of a sample. 
In this paper we apply 18 widely employed entropy estimators to Shannon measures useful to the software engineer: entropy, mutual information and conditional mutual information. Moreover, we investigate how the estimators are affected by two main influential factors: sample size and domain size. Our experiments range over a large set of randomly generated joint probability distributions and varying sample sizes, rather than choosing just one or two well known probability distributions as in previous investigations.

Our most important result is identifying that the Chao-Shen and Chao-Wang-Jost estimators stand out for consistently converging more quickly to the ground truth, regardless of domain size and regardless of the measure used. They also tend to outperform the others in terms of accuracy as sample sizes increase. This discovery enables a significant reduction in data collection effort without compromising performance.   
\end{abstract}


\maketitle

\section{Introduction}

Entropy, a fundamental measure of uncertainty and information content in random variables, serves as a versatile tool in diverse scientific disciplines, including software engineering. In cryptography and data compression, entropy is an indicator of redundancy in data~\cite{Auli2023}, assisting strong encryption key generation and more effective compression. In cyber attacks, conditional entropy evaluates the amount of information an attacker gains from the leaked data~\cite{Heusser2010, Mesecan2023}. Statistical dependence of variables in software is quantifiable via mutual information, with direct implications for feature selection tasks and causal analysis~\cite{Chen2018, Beraha2019}. Entropy is also used to improve testing efficiency~\cite{entropic} and to detect hallucinations in LLMs~\cite{Farquhar2024DetectingHI}.

Knowing a probability distribution completely means knowing the entropy completely. In numerous complex practical scenarios, however, the true distribution governing the data is, at best, partially understood. In such circumstances, entropy, and other measures derived from it, cannot be known precisely but must be estimated~\cite{Willink2012}.
In response to the burgeoning demand for robust estimation techniques, recent years have witnessed a surge in the development and implementation of entropy estimators, scattered across different programming languages and multiple libraries. 
Each estimator in this ever growing group comes with its assumptions and requirements. Many are built on highly specialised mathematical bases which are opaque to most users. How is a developer needing to use information theory to know which estimator to opt for, when many of their basic assumptions are neither known, understood, or even present? 

 The absence of a universally optimal estimator applicable across all types of data distributions poses a fundamental hurdle to this challenge~\cite{Paninski2003}. Moreover, estimator performance is inherently contingent upon data and sampling, with estimators sometimes approaching the real entropy value with arbitrary slowness as sample size grows in under-sampled regimes~\cite{Antos2001}.
 
In this study, we evaluate a number of well known entropy estimators in settings that remove them from their intended ``comfort zone'' and apply them in realistic situations more relevant to software developers. We examine how well each of these estimators behaves across a range of possible distributions, possible sample sizes, possible input spaces and, perhaps most importantly, across different Shannon measures. 
As well as entropy, we examine the behaviour of the biased estimators when calculating  mutual and conditional mutual information as these measures allow reasoning about data relationships in software, e.g. measuring leakage of secure information.

By systematically investigating the parameters above, we identify interesting patterns and insights that highlight the strengths and limitations of the estimators in different scenarios. As a result we can provide recommendations on which estimator to use even when little is known about the  underlying distributions. The contributions of this paper are:

\begin{enumerate}
    \item the first study of biased entropy estimators applied to Shannon measures other than entropy;
    \item a large experimental effort based on the study of 18 entropy estimators on datasets of twelve different lengths sampled from probability distributions generated on the basis of the following strategy: six new arbitrary probability distributions simulated for each of the three Shannon information quantities of interest (entropy, mutual information and conditional mutual information), each characterised by a different domain size, for a total of 1000 repetitions;
    \item and rigorous recommendations for which estimator(s) to use given unknown distributions and sample size adequacy. Our analysis identifies estimators that offer good accuracy under varying sample and domain sizes.
\end{enumerate}

The complete code implementation and evaluation outcomes presented in this paper are accessible at \url{https://figshare.com/s/24fae7566ca00662d538}.

\section{Preliminaries}

In this section, we provide a simple motivating example for the application of information theory in software engineering. We show that different estimators yield significantly different estimates from the same data. 

We then outline key uses of information theory and all necessary theoretical material. It is worth noting that we use the term ``entropy'' throughout the paper to refer explicitly to Shannon entropy. While Shannon entropy is just one member in the R\'enyi entropy family, it is the most commonly considered in the software engineering literature. Additionally, we do not address the problem of estimating differential entropy, which might be used in signal processing or electrical engineering. Programs and functions are fundamentally either discrete or discretised. 

\subsection{Motivating Example}

Consider the \lstinline{TriangleType} function~\cite{mesecan2021} in Listing \ref{lst:triangle}, which is used to determine the type of triangle based on the lengths of its three sides. For information flow analysis purposes, the first side is treated as a \emph{high security level} input. A software engineer may have several questions about the information flow properties of this function: How much information flows from input to output? How much secret information is revealed? What is the channel capacity of the function? If Listing \ref{lst:triangle} were the entire program, it would probably be feasible to test it relatively exhaustively. However, such a testing regime, even on this simple program, would still be expensive. If the function is embedded deep within the control flow of the software under test (SUT), it might not be possible to test the interface directly. It might also be impossible to control how many times the function is called. We need to estimate only on the basis of observed behaviors, and these observations may be sparse. This testing scenario is typically found in fuzzing~\cite{mesecan2021}.

\begin{lstlisting}[language=C, label={lst:triangle}, caption={A triangle program with information leakage}]
TriangleType typeOf(int high, int low1, int low2) {
    if (h == low1 && low1 == low2) {
        return EQULATERAL;
    } else if (high == low1 || high == low2 || low1 == low2) {
        return ISOSCELES;
    }
    return SCALENE;
}
\end{lstlisting}

The non-interference property\cite{Goguen1982} states that high-level information must not flow to low-level information (\emph{i.e.}, observable output). This implies that high-level information should not be involved in conditions or calculations from which the value of the outcome can be derived. It follows that the program under study clearly leaks secret information, as the \emph{high} variable is used as a boolean guard. In fact, the value of the output depends exclusively on the value taken on by the \lstinline{high} variable (\emph{e.g.}, if \lstinline{low1=2} and \lstinline{low2=3}, the triangle can be isosceles if \lstinline{high=2} or \lstinline{high=3}, and scalene otherwise).

To approximate the answer to the question of how much information is leaked, we calculate the Conditional Mutual Information (CMI) between the secret variable (\lstinline{high}) and the public (observable) output, conditioned on the low inputs. The simplest estimator, the \emph{Plugin} or \emph{Maximum Likelihood}, is just to count inputs and outputs. On just $500$ samples, the CMI of Listing \ref{lst:triangle} is very low: just $0.004$. With the exact same data, however, the Chao-Shen estimator gives $0.127$, a remarkable orders-of-magnitude difference. The Shrink estimator gives $0.003$, Bonachela $0.018$. Which number should the software engineer trust? It may have taken a long time even to generate $500$ samples and the exact nature of the input distribution may be unknown. 

In this paper, we ask which estimator, or estimators, work best on average when samples, testing flexibility, and distribution information are limited. Such situations are frequently encountered in software testing and security analysis. An analytic solution to this problem is unlikely to exist: each estimator comes with its own set of assumptions and limitations. Finding a ``best'' that works in all scenarios purely via analysis is unlikely to succeed. This is only reasonable: there is no perfect estimator. We answer the question then using empirical methods (Section \ref{sec:method}).

\subsection{Information Theory}

Information theory in software engineering is entering the mainstream.
Clark \etal~\cite{Clark2015} demonstrate how various aspects of software engineering can be improved by adopting an information-theoretical approach. Specifically, computer programs are essentially information transformation processes that adhere to fundamental mathematical limits, which can be rigorously studied using principles of information theory (IT). These insights provide the foundation for successful testing techniques, such as Entropic~\cite{entropic} and F-Bleau~\cite{Cherubin2019FBLEAUFB}. Outside of testing, IT is used in cryptography, data compression, information leakage, fault localisation, trustworthy AI and feature selection (see \Cref{sec:relwork}). In all of these areas, and more, it is rare to know the distributions underlying complex processes. Even when known, the computational cost of calculating exact measures can be high \cite{YasuokaTerauchiCSF2010}. This necessitates estimation. Entropic, for example, uses a simple Bayesian estimator, whereas Farquar \etal~\cite{Farquhar2024DetectingHI} use the plug-in estimator, albeit with a number of small, non statistical, modifications in both cases. Even with large sample sizes, the \emph{na\"ive} plug-in estimator (more commonly known as the maximum likelihood) does not fare well compared to estimations from more sophisticated methods.

\subsection{Shannon Measurements}

In information theory, ``the average level of information needed to describe the value of a random variable'' is expressed in terms of \emph{Shannon Entropy} $H(X)$~\cite{Shannon1948}. For a discrete random variable $X$ with a set of $k$ possible outcomes $\{x_1, x_2, \ldots, x_k\}$ (the alphabet or domain) and corresponding probabilities $\{p_1, p_2, \ldots, p_k\}$, the Shannon entropy is defined as:
\begin{equation}\label{eq-1}
    H(X) = - \sum_{x \in X} p(x) \log p(x)
\end{equation}
where the base of the $\log$ dictates the unit of information.
Entropy indicates the average uncertainty, or \emph{surprisal}, of $X$. The maximum entropy is achieved for $X$ when every element $x \in X$ is equally likely, corresponding to the uniform distribution~\cite{CoverThomas2006}. If the distribution is known, precise entropy calculation is straightforward. However, when the distribution of the observed data is unknown, the challenge of estimation arises. Estimation is particularly difficult when the sample size is insufficient to fully characterize the underlying distribution, often leading to poor estimations.

If we calculate a histogram of $N$ independent and identically distributed (\emph{i.i.d.}) samples drawn from $X$ and plug it directly into~\cref{eq-1}, we obtain:

\begin{equation}
\tag{1a}
\label{eq-1a}
 \hat{H}(X) = - \sum_{x \in X} (\hat{p}(x) \log \hat{p}(x))
\end{equation}

On average, this results in an underestimation of the true entropy~\cite{Basharin1959}. As discussed in the literature, the \emph{na\"ive} plug-in estimator $\hat{H}(X)$ (eq. (1a)) asymptotically approaches the true entropy $H(X)$ as $N \rightarrow \infty$ (assuming $k$ is fixed)~\cite{CoverThomas2006}. The larger the sample, the more closely the empirical distribution $\hat{p}(x)$ approximates the true probability distribution $p(x)$.

The empirical probability $\hat{p}(x)$ is given by $\hat{p}(x) = \frac{\mathrm{n}_{x}}{N}$, where $\mathrm{n}_{x}$ is the number of occurrences of the value $x$ in the sample. The expected systematic error (bias) $\Delta = E[\hat{H}(X)] - H(X)$ of the estimator increases for small $N$, until $N \gg k$. More precisely, considering the approximation error $\epsilon_{x} = \frac{\hat{p}(x) - p(x)}{p(x)}$ and using some algebraic expansions, it is possible to express the derived entropy as $\hat{H}(X) = H(X) - \frac{m-1}{2N}$~\cite{Roulston1999}, where $m$ denotes the number of non-zero probability values. This demonstrates the systematic downward bias in estimating the entropy due to finite sample size, introduced by the term $-\frac{m-1}{2N}$. Bounds on such bias are given by:

\[
-\log\left(1 + \frac{m-1}{N}\right) \leq \Delta \leq 0
\]

This relationship leverages the expectation $E[\hat{H}(X)] - H(X) = -E(D_{KL}(\hat{p} \parallel p))$, Jensen's inequality, and the properties of Kullback-Leibler (KL) divergence~\cite{Paninski2003}.

A considerable body of research has focused on implementing methods to mitigate the inherent systematic error of entropy estimators. Miller~\cite{Miller1955} formulated the first bias-corrected version of the na\"ive estimator, paving the way for a series of ever more sophisticated approaches aimed at improving the reliability of entropy estimation. Notable advancements include the Chao-Shen sample coverage-based adjustment~\cite{ChaoShen2003} and the NSB correction via prior distribution integration~\cite{Nemenman2002}, among others. 

\textit{Mutual Information $I(X;Y)$}~\cite{Shannon1948} is defined as ``the amount of information conveyed, on average, by one random variable $X$ about another $Y$'', i.e., the amount of shared information. Less formally, it measures the reduction in uncertainty of a random variable $Y$ due to the knowledge of another random variable $X$, or \emph{vice versa}.
\textit{Conditional Mutual Information $I(X;Y|Z)$}~\cite{Shannon1948} refers to the information shared between two variables ($X$ and $Y$) when a third variable ($Z$) is known.
The mathematical representations for both the mutual information (MI) and the conditional mutual information (CMI) can reduce to entropy-based formulations. Importantly, in a multivariate setting, $X$, $Y$, and $Z$ can refer to joint random variables, rather than individual random variables.

In this study, we examine discrete entropy estimators only, as these commonly represent data in conventional software. Further research is required to evaluate estimators for continuous random variables. Entropy estimators can be broadly classified into two categories: Frequentist and Bayesian~\cite{Willink2012}. Both offer unique frameworks for data analysis, rooted in different philosophical interpretations of probability and statistical reasoning. Frequentist entropy estimators typically rely on observed frequencies of events to estimate entropy. For example, the plug-in estimator calculates entropy directly from the empirical distribution of the sample.
The Bayesian approach, on the other hand, interprets probability as a measure of belief or certainty about an event, which can be updated as new evidence is acquired. This method explicitly incorporates prior information through the use of prior distributions, which are combined with the likelihood of the observed data to produce posterior distributions. Therefore, Bayesian entropy estimators incorporate prior distributions over the probability distribution that is being estimated. This can help mitigate biases introduced by small sample sizes and improve robustness.





\section{Estimating more than Entropy with Entropy Estimators}\label{sec:method}

We consider the problem of estimating Shannon measurements from discrete data, in cases where the probability distribution is not known \emph{a priori}.

We experimentally evaluate the performance of widely recognised entropy estimators in contexts that differ from those in which they have been characterized by previous comparative studies. Prior research has focused on narrow experimental settings constrained by particular distribution classes and discretization techniques (see~\Cref{sec:relwork}). We ensure that none of these influence our observations; rather, we isolate and analyze the effects of two main influential factors: sample size and domain size. By systematically varying these parameters, we conduct a generalized performance analysis that can inform a wide range of real-world applications under different data availability scenarios. All evaluations are conducted with simulated data within the framework of general probability distributions, as described in~\Cref{subsec:design}.



We offer a clear understanding of which estimator(s) show higher accuracy in the well-sampled and under-sampled regimes, where estimation can be severely biased~\cite{Treves1995, Montalvao2012, Macke2013, Hernández2022}. 
Specifically, we extend our investigation to mutual information (MI) and conditional mutual information (CMI) too. At the same time, we study the degree of  dependence of the estimator to sample size by analyzing the convergence speed to the ground truth with respect to growing samples and domain sizes. The convergence point is used to inform the ratio (sample size to domain size) at which estimator performance becomes stable and independent of the sample size. Therefore, our study allows us to outline selective criteria for strengths and weaknesses, along with assessing potential behavioral stability across different circumstances. This way, we seek to equip practitioners with the knowledge needed to make informed decisions on which estimator to employ and how to utilize it effectively in their specific scenarios, even in the face of limited knowledge of the underlying distributions.

We examine 18 popular discrete entropy estimators, implemented in the Julia DiscreteEntropy package~\cite{kelly2024discreteentropy}, as detailed in~\Cref{tab:h_estimators}. Using implementations all from one package has two distinct advantages: simplicity and, more importantly, uniformity of precision. It is difficult to compare estimators drawn from different programming languages when each implementation has a differing default number of significant digits, and the estimates themselves often vary by small amounts. To the best of our knowledge, no previous study has taken into account the effect of varying significance levels when comparing estimators from different packages. 

\begin{table}[t!]
\centering
    \begin{tabular}{p{0.7mm} | l | c | c |}
    \cline{2-4}
        & 
        \multicolumn{1}{l|}{\textbf{Estimator}} &
        \multicolumn{1}{c|}{\textbf{(\textit{est})}} & \multicolumn{1}{l|}{\textbf{Source}} \\
        \Xcline{2-4}{2\arrayrulewidth}
        & MaximumLikelihood & ML & \cite{Paninski2003}\\
        \cline{2-4}
        & MillerMadow & MM & \cite{Miller1955} \\
        \cline{2-4}
        & Grassberger88 & GSB\tiny{88} & \cite{Grassberger1988} \\
        \cline{2-4}
        & Grassberger03 & GSB\tiny{03} & \cite{Grassberger2003} \\
        \cline{2-4}
        & Schurmann & SHU & \cite{Schurmann2004} \\
        \cline{2-4}
        & Chao-Shen & CS & \cite{ChaoShen2003} \\
        \cline{2-4}
        & Zhang & Z & \cite{Zhang2012} \\
        \cline{2-4}
        & Shrink & SHR & \cite{Hausser2009} \\ 
        \cline{2-4}
        & Bonachela & B & \cite{Bonachela2008} \\
        \cline{2-4}
        & Chao-Wang-Jost & CW & \cite{ChaoWang2013} \\
        \Xcline{2-4}{2\arrayrulewidth}
        & Pym & PYM & \cite{JMLR:v15:archer14a}\\ 
        \cline{2-4}
        & Bayes & BAY & \cite{rubin1981bayesian}\\
        \cline{2-4}
        & LaPlace & LAP & \cite{Holste1999_LaPlace} \\
        \cline{2-4}
        & Jeffrey & JEF & \cite{Krichevsky1981_Jeffrey} \\
        \cline{2-4}
        & SchurmannGrassberger & SG & \cite{Grassberger2021_generalized} \\
        \cline{2-4}
        & Minimax & MIN & \cite{Stanislaw1958_MinMax} \\
        \cline{2-4}
        & NSB & NSB & \cite{Nemenman2002} \\
        \cline{2-4}
        & ANSB & ANSB & \cite{Nemenman2004} \\
        \cline{2-4}
    \end{tabular}
    \caption{Biased entropy estimators analysed in this work. A list of abbreviations is provided in the second column (\emph{est}) for quick reference.}
    \label{tab:h_estimators}
\end{table} 

\subsection{Research Questions}
This study answers the following three research questions:

\textit{\textbf{RQ1 - Convergence:} Which estimators stabilise most quickly when estimating H, MI and CMI for an unknown probability distribution?}

This research question investigates how the estimates for entropy (H), mutual information (MI), and conditional mutual information (CMI) vary across samples of increasing size. We detect the sample size at which the performance of a given estimator stabilises within a predefined \emph{flattening-off bound}, indicating negligible change in estimation accuracy with additional data. We focus on estimators that stabilise with fewer samples, i.e. that achieve reliable estimations using the least amount of sample data. The smaller this ``sufficient'' (or ``safe'') sample size, the faster the convergence to the ground truth. This is particularly valuable in scenarios where data collection is resource-intensive or limited.

\textit{\textbf{RQ2 - Sample size:} What is the point where the optimal estimators found in RQ1 stabilise?}

The answer to RQ1 reveals the fastest converging estimators. RQ2 provides deeper insight into their convergence points, i.e. the minimal number of samples required to achieve consistent and reliable outcomes. Essentially, we explore the ratios of the ``safe'' sample size to domain size, for growing domain sizes, to uncover systematic patterns or logical relationships in the data. 
This provides a clear picture of how sample size requirements scale with the complexity or diversity of the input space. Therefore, it offers valuable indications to systematically determine the extent to which data collection can be minimised without compromising the estimation accuracy, according to specific requirements.

\textit{\textbf{RQ3 - Accuracy:} Which are the most accurate estimators when estimating H, MI and CMI for an unknown probability distribution?}

We compare the performance of the estimators in terms of convergence speed and accurate estimate of the ground truth, aiming to identify potential correlations.




\subsection{Experimental Design}\label{subsec:design}
We examine discrete entropy estimators, extending their application to the measurement of mutual information and conditional mutual information, which prove to be more insightful in complex software engineering applications. Specifically, our investigation centers around understanding how their performance is affected by the sample size and domain size, when sampling from general probability distributions (PDs). By doing so, we seek the estimators that perform best on average across various sampling scenarios and measures. That with the aim of providing software engineers with robust recommendations applicable to a wider range of real applications than those assisted by entropy measurements only.

\textit{Baseline Entropy Estimators} - We select 18 most widespread discrete entropy estimators from the literature. A comprehensive illustration is presented in \Cref{tab:h_estimators}. As clearly shown, these estimators are categorised into two classes: Bayesian estimators, such as NSB~\cite{Nemenman2002} and
PYM~\cite{pym2014}, and Frequentist estimators, such as Grassberger~\cite{grassberger2004}. 

We thus exploit the joint entropy-based formulation of the mutual and conditional mutual information to derive our MI and CMI estimators, as outlined in~\cref{eq:mi,eq:cmi}. As the estimators in~\Cref{tab:h_estimators} are all \emph{entropy} estimators, we use  
\begin{equation}
    \hat{I}_{est}(X;Y) = \hat{H}_{est}(X) + \hat{H}_{est}(Y) - \hat{H}_{est}(X,Y)
    \label{eq:mi}
\end{equation}
for MI and
\begin{equation}
\begin{split}
    \hat{I}_{est}(X;Y|Z) = \hat{H}_{est}(X,Z) + \hat{H}_{est}(X,Y,Z)  \\
    - \hat{H}_{est}(Y,Z) - \hat{H}_{est}(Z)
\end{split} 
\label{eq:cmi}
\end{equation}
for CMI. In both cases, $\hat{H}_{est}$ is one of the 18 estimators found in~\Cref{tab:h_estimators}.

We implemented everything in the Julia language\footnotemark[\value{footnote}] and take $\log$ of our information-theoretic implementations to be the natural logarithm, giving entropy in $nats$.

In line with the questions posed by this study, we designed two main experiments aimed at evaluating respectively the convergence speed (RQ1-2) and the accuracy (RQ3) of the estimators in different analytical contexts.

\subsubsection{Data Collection}

To analyse the performance of the entropy estimators w.r.t. entropy, mutual and conditional mutual information, we first assumed knowledge on the real nature (underlying probability distribution) of the observed data. This way, we were able to access the ground truth (actual value) of the information-theoretic measures of interest, crucial for validating the effectiveness of the different estimation methods under analysis. To that end, we randomly generated respectively 1D, 2D and 3D arrays to simulate $n$-variate discrete distributions on a $k$-ary alphabet (i.e. arbitrary probability mass functions $f:\mathbb{R}^n\rightarrow[0,1]\Rightarrow f(\mathrm{x})=p_\mathrm{x},$ $ \forall n\in\{1,2,3\} \, \vert \, \mathrm{x}\in \{(x),(x,y),(x,y,z)\} \land x,y,z\in X,Y,Z$). To make it clear: $n$ denotes the number of discrete random variables (or sets of discrete random variables) involved into the analysis of the H, MI and CMI estimations, respectively one ($X$), two ($X,Y$) and three ($X,Y,Z$); $k$ represents the domain size of the corresponding probability distribution (to be meant as ``joint probability distribution'' in case of MI and CMI). In a nutshell, $k$ refers to the space in which a random variable following that distribution takes its values, i.e. the set of points\footnote{All possible combinations of values for the $n$ variables} in $\mathbb{R}^n$ where $f$ is non-zero ($\text{supp}(f)=\{\mathrm{x}\in \mathbb{R}^n|f(\mathrm{x})\neq 0\}$). 
The following equation shows the structure of the contingency matrices used to model arbitrary joint probability distributions describing the nature of the observed (sampled) data used for our estimations.
  \[
  \text{PD\textsubscript{$\mkern-3muX\mkern-3muY$}} = 
  \begin{bmatrix}

  p_{x_1y_1} {\text{(1)}} & p_{x_1y_2} & \ldots & p_{x_1y_\mathrm{m}} \\
  p_{x_2y_1} & p_{x_2y_2} & \ldots & p_{x_2y_\mathrm{m}} \\
  \vdots & \vdots & \ddots & \vdots \\
  p_{x_\mathrm{n}y_1} & p_{x_\mathrm{n}y_2} & \ldots & p_{x_\textbf{n}y_\textbf{m}} {\text{($k$)}}\\
  \end{bmatrix} : 
  \begin{cases}
    k = \mathrm{n} \times \mathrm{m}
  \end{cases}
   \;  
  \]

The equation shows the bi-dimensional scenario (two random variables $X$ and $Y$ involved in MI measurements), but it applies also to CMI by moving to a three-dimensional space. The $k$ entries in the matrix correspond to the probabilities of the domain, i.e. the possible combinations of outcomes from the discrete variables involved.



After simulating categorical distributions (``join'' for MI and CMI) of varying dimensions, we sampled from them to generate observed data useful for our estimations. Specifically, we executed multiple samplings from a given distribution, by varying the number of individual observations ($N$) drawn each time. By doing so, we aimed to understand how different sample sizes affect the accuracy and reliability of the estimations. To ensure robustness and mitigate the impact of randomness, the entire process was repeated 1000 times (a number deemed sufficiently high based on established practices in similar studies). This means that, for a given $k$, we simulated $r=$ 1000 random distributions and, from each of them, we systematically extracted samples of varying size. 

It is worth noting that we considered different distribution configurations also according to the value of the ground truth. The aim was to capture the full spectrum of possible information content of the data and ensure that our analysis comprehensively covers from low to high levels of entropy (H), mutual information (MI), and conditional mutual information (CMI).

\begin{figure}[t]
\begin{minipage}{\columnwidth}
\end{minipage}

\caption{}
\label{fig:contingency_matrices}
\end{figure}

\begin{table} [t]
    \centering
    \begin{adjustbox}{width=0.65\columnwidth}
    \begin{tabular}{ c c c | l }
    \hline
    H & MI & CMI &  \\
    \hline
    \multirow{2}{*}{
    PD\textsubscript{$\mkern-3muX$}($gt, k, r$)} & \multirow{2}{*}{PD\textsubscript{$\mkern-3muX\mkern-3mu Y$}($gt, k, r$)} & \multirow{2}{*}{PD\textsubscript{$\mkern-3muX\mkern-3muY\mkern-3muZ$}($gt, k, r$)} & $gt\in$\{${S}, M,L$\}\\
    & & & $k=2^i, i\,\shorteq\,8..18$\\
    \hline
    $S$\textsubscript{$\mkern-3muX$}($N, r$) & $S$\textsubscript{$\mkern-3muX\mkern-3muY$}($N, r$) & $S$\textsubscript{$\mkern-3muX\mkern-3muY\mkern-3muZ$}($N, r$) & $N=2^i, i\,\shorteq\,3..14$\\
    \hline
    \\
    \multicolumn{4}{l}{
    $
    S_\mathrm{\text{X}}^r = [s_1, s_2, \ldots, N] \sim \text{PD}_\mathrm{\text{X}}^r, \quad \forall
    \begin{cases} 
    r = 1..1000
     \\
    \mathrm{X} \in \{X, XY, XYZ\}
    \end{cases}$}\\ \\
    
    \end{tabular}
    \end{adjustbox}
    \caption{Overview of the experimental settings. The first row in the table shows the possible configurations of the simulated discrete probability distributions (PD), defined by the level of ground truth ($gt$) and the size of the domain ($k$). For any specific $gt$ scenario and Shannon information quantity (H, MI or CMI), a new set of 1000 ($r$) arbitrary distribution is generated six times, each with a different domain size ($k$). The second row shows that, from each of the so generated PDs, datasets of variable length ($N$) are sampled. }
    \label{tab:exp_setup}

\end{table}

\subsubsection{Experimental Setup}

The configuration of our experimentation can be delineated along three main distinct standpoints (\Cref{tab:exp_setup}):
\begin{itemize}
    \item Probability distribution (PD) domain size: we gradually increased the size of the input space\footnote{($\mathrm{n}$), ($\mathrm{n} \times \mathrm{m}$) or ($\mathrm{n} \times \mathrm{m} \times \mathrm{l}$) where $\mathrm{n, m, l}$ are the domain sizes of the variables $X, Y, Z$.} of the modelled PDs, considering six configurations of the domain, with $k=$ 256, 1024, 4096, 16384, 65536, 262144;
    \item Sample size: we sampled finite data sets of growing length $N=$ 8, 16, 32, 64, 128, 256, 512, 1024, 2048, 4096, 8192, 16384 (varying in the interval [$2^{3}-2^{14}$]); 
    \item Ground truth: the usage of a completely uniform random approach to populate large joint probability distributions would have come out to show information content and statistical dependency between the variables $X$, $Y$ and $Z$ close to zero every time. Therefore, motivated by the necessity to ensure some level of statistical dependency allowing us to test the entropy estimators under quasi-random distributions, we generated probability distributions by considering three ground truth scenarios:
    \begin{itemize}
    \item [-]\emph{small (S)} H, MI, CMI: the variables exhibit information content and interdependence close to the minimal $\rightarrow H_{min}(X), I_{min}(X; Y), I_{min}(X; Y| Z) = 0 $
    \item [-]\emph{medium (M)} H, MI, CMI: unstructured scenarios
    \item [-]\emph{large (L)} H, MI, CMI: the variables exhibit levels of uncertainty and interdependence close to the maximum $\rightarrow H_{max}(X)=\log (k),
    I_{max}(X;Y)=\min(H(X),\\H(Y)),I_{max}(X;Y|Z)=\min(H(X|Z),H(Y|Z))$

     \end{itemize} 
\end{itemize}

We choose \emph{power-of-two} sample sizes to allow easy comparisons with findings from other pertinent studies using the same sampling approach (like the work by Rodriguez \etal~\cite{Rodriguez2021}). 

All the experiments \hide{for each configuration)}were run on a Linux Debian machine (12 core, 24 threads, 128GB RAM). An overall overview of the experimental settings is illustrated in~\Cref{tab:exp_setup}.

\section{Results}\label{sec:results}

Our analysis has been developed on averaged data (estimations) resulting from multiple runs (1000). We have examined the performance of 18 entropy estimators with respect to the entropy, mutual information, and conditional mutual information, according to different characterizations of the influential parameters of interest. Specifically, within a given ground truth ($gt$) scenario, we fixed the domain size ($k$) while varying the sample size ($N$). Essentially, we studied both the accuracy and convergence speed of the estimations obtained by using each estimation method with samples of increasing size in the interval [$2^{3}$-$2^{14}$].

The following is a comprehensive description of the experimental process from different perspectives of data interpretation. This aims at drawing interesting conclusions for varying hypothetical operational scenarios. The results refer to the mean of the estimated measures across the different levels of $gt$ considered in this work. This approach helps in understanding the overall trends and patterns of the estimations without assuming any knowledge about the ground truth. It provides a more objective evaluation of the estimators' performance.

\newcolumntype{N}{>{\mdseries}^w{c}{2mm}}
\begin{table*}[t!]
\begin{tabular}{c c c}
    \begin{subtable}{0.31\textwidth}
    \begin{adjustbox}{width=\textwidth}
    \begin{tabular}{ l *{6}{N} : l *{2}{N}}
        \hline
        \rowstyle{} \diagbox{\emph{est}}{\emph{k}} & \rotatebox{60}{256} & \rotatebox{60}{1024} & \rotatebox{60}{4096} & \rotatebox{60}{16384} & \rotatebox{60}{65536} & \rotatebox{60}{262K} & \rotatebox{90}{slope} & \rotatebox{90}{$l_2$-norm} \\
        \hline
        ML & \cellcolor{gray!25}9 & \cellcolor{gray!25}10 & \cellcolor{gray!25}12 & & & & 1.5 & 31.6 \\
        MM & \cellcolor{ao(english)!35}9 & \cellcolor{ao(english)!35}9 & \cellcolor{ao(english)!35}11 & \cellcolor{ao(english)!35}13 & & & 1.33 & 30.0 \\
        GSB\tiny{88} & \cellcolor{ao(english)!35}9 & \cellcolor{ao(english)!35}9 & \cellcolor{ao(english)!35}11 & \cellcolor{ao(english)!35}13 & & & 1.33 & 30.0 \\
        GSB\tiny{03} & \cellcolor{ao(english)!35}9 & \cellcolor{ao(english)!35}9 & \cellcolor{ao(english)!35}11 & \cellcolor{ao(english)!35}13 & & & 1.33 & 30.0 \\
        SHU & \cellcolor{ao(english)!35}9 & \cellcolor{ao(english)!35}9 & \cellcolor{ao(english)!35}11 & \cellcolor{ao(english)!35}13 & & & 1.33 & 30.0 \\
        \textbf{CS} & \cellcolor{bluebell!60}\textbf{9} & \cellcolor{bluebell!60}\textbf{9} & \cellcolor{bluebell!60}\textbf{10} & \cellcolor{bluebell!60}\textbf{9} & \cellcolor{bluebell!60}\textbf{10} & \cellcolor{bluebell!60}\textbf{12} & \textbf{0.5} & \textbf{24.23} \\
        Z & \cellcolor{ao(english)!35}9 & \cellcolor{ao(english)!35}9 & \cellcolor{ao(english)!35}11 & \cellcolor{ao(english)!35}13 & & & 1.33 & 30.0 \\
        SHR & \cellcolor{gray!25}9 & \cellcolor{gray!25}10 & \cellcolor{gray!25}12 & & & & 1.5 & 31.6 \\
        B & 11 & 13 & & & & & & \\
        \textbf{CW} & \cellcolor{bluebell!60}\textbf{9} & \cellcolor{bluebell!60}\textbf{9} & \cellcolor{bluebell!60}\textbf{10} & \cellcolor{bluebell!60}\textbf{9} & \cellcolor{bluebell!60}\textbf{10} & \cellcolor{bluebell!60}\textbf{12} & \textbf{0.5} & \textbf{24.23} \\
        PYM & & & & & & & & \\
        BAY & \cellcolor{gray!25}9 & \cellcolor{gray!25}10 & \cellcolor{gray!25}12 & & & & 1.5 & 31.6 \\
        LAP & \cellcolor{gray!25}9 & \cellcolor{gray!25}10 & \cellcolor{gray!25}12 & & & & 1.5 & 31.6 \\
        J & \cellcolor{gray!25}9 & \cellcolor{gray!25}10 & \cellcolor{gray!25}12 & & & & 1.5 & 31.6 \\
        SG & \cellcolor{gray!25}10 & \cellcolor{gray!25}10 & \cellcolor{gray!25}12 & & & & 1.5 & 31.6 \\
        MIN & & & & 11 & 12 & 12 & & \\
        NSB & & & & & & & & \\
        ANSB & & & & & & & & \\
        \hline
    \end{tabular}
    \end{adjustbox}
    \caption{H}
    \label{tab:Fp_tab_h}
    \end{subtable} &
    \begin{subtable}{0.31\textwidth}
    \begin{adjustbox}{width=\textwidth}
    \begin{tabular}{ l *{6}{N} : l *{2}{N} }
        \hline
        \rowstyle{} \diagbox{\emph{est}}{\emph{k}} & \rotatebox{60}{256} & \rotatebox{60}{1024} & \rotatebox{60}{4096} & \rotatebox{60}{16384} & \rotatebox{60}{65536} & \rotatebox{60}{262K} & \rotatebox{90}{slope} & \rotatebox{90}{$l_2$-norm} \\
        \hline
        ML & \cellcolor{gray!25}8 & \cellcolor{gray!25}11 & \cellcolor{gray!25}13 & & & & 2.5 & 32.1 \\
        MM & \cellcolor{gray!45}8 & \cellcolor{gray!45}10 & \cellcolor{gray!45}12 & & & & 2.0 & 31.4 \\
        \textbf{GSB\tiny{88}} & \cellcolor{ao(english)!35}\textbf{7} & \cellcolor{ao(english)!35}\textbf{9} & \cellcolor{ao(english)!35}\textbf{11} & \cellcolor{ao(english)!35}\textbf{13} & & & \textbf{2.0} & \textbf{29.5} \\
        GSB\tiny{03} & \cellcolor{gray!45}8 & \cellcolor{gray!45}10 & \cellcolor{gray!45}12 & & & & 2.0 & 31.4 \\
        \textbf{SHU} & \cellcolor{ao(english)!35}\textbf{7} & \cellcolor{ao(english)!35}\textbf{9} & \cellcolor{ao(english)!35}\textbf{11} & \cellcolor{ao(english)!35}\textbf{13} & & & \textbf{2.0} & \textbf{29.5} \\
        \textbf{CS} & \cellcolor{bluebell!60}\textbf{6} & \cellcolor{bluebell!60}\textbf{8} & \cellcolor{bluebell!60}\textbf{9} & \cellcolor{bluebell!60}\textbf{11} &  & & \textbf{1.9} & \textbf{27.4} \\
        Z & \cellcolor{gray!45}8 & \cellcolor{gray!45}10 & \cellcolor{gray!45}12 & & & & 2.0 & 31.4 \\
        SHR & \cellcolor{gray!45}8 & \cellcolor{gray!45}10 & \cellcolor{gray!45}12 & & & & 2.0 & 31.4 \\
        B & 11 & & & & & & & \\
        \textbf{CW} & \cellcolor{bluebell!90}\textbf{6} & \cellcolor{bluebell!90}\textbf{7} & \cellcolor{bluebell!90}\textbf{9} & \cellcolor{bluebell!90}\textbf{10} & \cellcolor{bluebell!90}\textbf{12} & \cellcolor{bluebell!90}\textbf{13} & \textbf{1.46} & \textbf{24.1} \\
        PYM & & & & & & & & \\
        BAY & \cellcolor{gray!25}8 & \cellcolor{gray!25}11 & \cellcolor{gray!25}13 & & & & 2.5 & 32.1 \\
        LAP & \cellcolor{gray!45}8 & \cellcolor{gray!45}10 & \cellcolor{gray!45}12 & & & & 2.0 & 31.4 \\
        J & \cellcolor{gray!45}8 & \cellcolor{gray!45}10 & \cellcolor{gray!45}12 & & & & 2.0 & 31.4 \\
        SG & \cellcolor{gray!25}8 & \cellcolor{gray!25}11 & \cellcolor{gray!25}13 & & & & 2.5 & 32.1 \\
        MIN & \cellcolor{gray!25}8 & \cellcolor{gray!25}11 & \cellcolor{gray!25}13 & & & & 2.5 & 32.1 \\
        NSB & \cellcolor{gray!45}8 & \cellcolor{gray!45}10 & \cellcolor{gray!45}12 & & & & 2.0 & 31.4 \\
        ANSB & & & & & & & & \\
        \hline
    \end{tabular}
    \end{adjustbox}
    \caption{MI}
    \label{tab:Fp_tab_mi}
    \end{subtable} &
    \begin{subtable}{0.302\textwidth}
    \begin{adjustbox}{width=\textwidth}
    \begin{tabular}{ l *{6}{N} : l *{2}{N} }
        \hline
        \rowstyle{} \diagbox{\emph{est}}{\emph{k}} & \rotatebox{60}{256} & \rotatebox{60}{1024} & \rotatebox{60}{4096} & \rotatebox{60}{16384} & \rotatebox{60}{65536} & \rotatebox{60}{262K} & \rotatebox{90}{slope} & \rotatebox{90}{$l_2$-norm} \\
        \hline
        ML & \cellcolor{gray!60}4 & \cellcolor{gray!60}10 & \cellcolor{gray!60}12 & & & & 4.0 & 30.6 \\
        MM & \cellcolor{gray!45}7 & \cellcolor{gray!45}10 & \cellcolor{gray!45}12 & & & & 2.5 & 31.1 \\
        \textbf{GSB\tiny{88}} & \cellcolor{ao(english)!55}\textbf{6} & \cellcolor{ao(english)!55}\textbf{9} & \cellcolor{ao(english)!55}\textbf{11} & \cellcolor{ao(english)!55}\textbf{13} & & & \textbf{2.3} & \textbf{29.3} \\
        GSB\tiny{03} & \cellcolor{gray!80}7 & \cellcolor{gray!80}9 & \cellcolor{gray!80}12 & & & & 2.5 & 30.8 \\
        \textbf{SHU} & \cellcolor{ao(english)!35}\textbf{7} & \cellcolor{ao(english)!35}\textbf{9} & \cellcolor{ao(english)!35}\textbf{11} & \cellcolor{ao(english)!35}\textbf{13} & & & \textbf{2.0} & \textbf{29.5} \\
        \textbf{CS} & \cellcolor{bluebell!60}\textbf{6} & \cellcolor{bluebell!60}\textbf{7} & \cellcolor{bluebell!60}\textbf{10} & \cellcolor{bluebell!60}\textbf{12} &  &  & \textbf{2.1} & \textbf{27.9} \\
        Z & \cellcolor{gray!80}7 & \cellcolor{gray!80}9 & \cellcolor{gray!80}12 & & & & 2.5 & 30.8 \\
        SHR & \cellcolor{gray!25}8 & \cellcolor{gray!25}10 & \cellcolor{gray!25}12 & & & & 2.0 & 31.4 \\
        B & 10 & & & & & & & \\
        \textbf{CW} & \cellcolor{bluebell!90}\textbf{6} & \cellcolor{bluebell!90}\textbf{7} & \cellcolor{bluebell!90}\textbf{8} & \cellcolor{bluebell!90}\textbf{13} & \cellcolor{bluebell!90}\textbf{12} & & \textbf{1.8} & \textbf{26.2} \\
        PYM & & & & & & & & \\
        BAY & \cellcolor{gray!60}4 & \cellcolor{gray!60}10 & \cellcolor{gray!60}12 & & & & 4.0 & 30.6 \\
        LAP & \cellcolor{gray!45}7 & \cellcolor{gray!45}10 & \cellcolor{gray!45}12 & & & & 2.5 & 31.1 \\
        J & \cellcolor{gray!45}7 & \cellcolor{gray!45}10 & \cellcolor{gray!45}12 & & & & 2.5 & 31.1 \\
        SG & \cellcolor{gray!60}4 & \cellcolor{gray!60}10 & \cellcolor{gray!60}12 & & & & 4.0 & 30.6 \\
        MIN & \cellcolor{gray!60}4 & \cellcolor{gray!60}10 & \cellcolor{gray!60}12 & & & & 4.0 & 30.6 \\
        NSB & \cellcolor{gray!25}8 & \cellcolor{gray!25}10 & \cellcolor{gray!25}12 & & & & 2.0 & 31.4 \\
        ANSB & 7 & 9 & & & & & & \\
        \hline
    \end{tabular}
    \end{adjustbox}
    \caption{CMI}
    \label{tab:Fp_tab_cmi}
    \end{subtable} 
    \end{tabular}
    \caption{Tabular illustration of the convergence stability for the 18 estimators of (a) entropy, (b) mutual information, and (c) conditional mutual information. The integer values in the tables represent the \emph{flattening-off points} (\emph{Fp}) of the estimations, observed across samples of increasing size. For visual clarity, these values refer to the exponent in the expression $2^{\text{exp}}$ (in line with the exponential notation we used to denote sample sizes). For example, the first entry in table (c) is $4$, indicating that the estimations obtained using the Maximum Likelihood (ML) method (for domains of size 256) exhibit negligible fluctuations ($<0.1$) for samples larger than $2^{4}$ (16). Rows with equal/similar colors reflect common/quasi-common flattening-off trends of the estimators over different domain sizes $k$. The fastest converging estimators are highlighted in bold (purple), with the Chao-Wang-Jost (CW) method showing a slight edge. This observation is confirmed by the slope and Euclidean distance measurements.}
    \label{tab:Fp_tab}
\end{table*}

\subsection{Convergence}




We investigate the stability of the estimations against the sample size. To that aim, for each estimator, we initially performed a delta analysis of the respective estimations for samples of increasing size. Specifically, we used the estimations' mean squared error (MSE) to conduct evaluations more sensitive to potential deviations and outliers in the data resulting from the 1000 runs. Delta ($\delta$) is formally defined as the overall change between two measurements. In the context of this study, the two measurements correspond to estimates calculated from two samples of increasing size (within the range under analysis). Therefore, our delta is the difference in the estimation as the sample size grows.
We consider this difference to be negligible when it falls below 0.1, which we refer to as the \emph{flattening-off bound (Fb)}. We believe such a difference represents an acceptable trade-off between ensuring minimal impact on the estimation quality and accounting for the inherent bias an estimation can be subject to. 

For instance, let's assume that, according to a certain entropy estimator, the mutual information of a sample $a$ with 256 elements and a sample $b$ with 1024 elements is respectively $\hat{I}_a=0.15$ and $\hat{I}_b=0.18$. Since we work with synthetic data, we can measure the deviation of these estimations from the actual MI value (e.g., $gt=0.12$). For the example above, the estimation error is $err\hat{I}_a=0.15-0.12=0.03$ and $err\hat{I}_b=0.18-0.12=0.06$. If the absolute difference between these estimation errors ($\delta=|err\hat{I}_a-err\hat{I}_b|$) is smaller than 0.1, we can claim that the performance of the estimator is stable against the two analysed sample sizes. This means that, within our accepted margin of fluctuation (\emph{Fb}), the behaviour of that estimator is not significantly affected by the sample size. In essence, the estimate remains relatively unchanged regardless of the amount of data observed.

We thus pinpoint the sample size at which the performance of a given estimator starts to stabilize under our established threshold (i.e., it does not fluctuate beyond the flattening-off bound). We define this sample size as the ``sufficient'' (or ``safe'') size for that estimator to deliver reliable estimations, or more formally, the \emph{flattening-off point (Fp)}. It is worth noting that this stability must be consistently observed for all samples with size (within our predefined range) larger than \emph{Fp}. This means that the result obtained by adopting that estimation method is not affected by the sample size if considering samples larger than \emph{Fp}. 

A direct benefit of this observation could be witnessed in cases where only small datasets are available, provided their size falls within the range marked by \emph{Fp}. In such cases, we expect the estimation's accuracy not to improve significantly ($<$0.1) with bigger samples. We use the safe sample size as an indicator of the convergence speed: the smaller it is, the faster the convergence to the ground truth is.

\begin{table*}[t]
    \centering
    \begin{adjustbox}{width=1\textwidth}
    \begin{tabular}{ l c c c c c c c c c c c c }
        \hline
        \multirow{2}{*}{\textbf{Estimators}} & \multicolumn{12}{c}{\textbf{Sample Size}} \\
        \cline{2-13} 
        & \textbf{$2^{3}$} & \textbf{$2^{4}$} & \textbf{$2^{5}$} & \textbf{$2^{6}$} & \textbf{$2^{7}$} & \textbf{$2^{8}$} & \textbf{$2^{9}$} & \textbf{$2^{10}$} & \textbf{$2^{11}$} & \textbf{$2^{12}$} &\textbf{$2^{13}$} & \textbf{$2^{14}$} \\ 
        \hline
        \textbf{Frequentist} & & & & & & & & & & & & \\

        ML
        & 1.0186 & 1.1612 & 1.0402 & 0.6740 & 0.3274 & \textbf{0.1241} & 0.0370 & 0.0105 & 0.0030 & 0.0009 & 0.0003 & 0.0001\\

        MM & 1.3316 & 1.3891 & 1.0185 & 0.5012 & 0.1731 & \textbf{0.0436} & 0.0081 & 0.002 & 0.0007 & 0.0003 & 0.0001 & 0.0001\\

        GSB\tiny{88} & 1.6342 & 1.1437 & 0.6271 & 0.2482 & \textbf{0.0728} & 0.0179 & 0.0042 & 0.0015 & 0.0006 & 0.0003 & 0.0001 & 0.0001\\

        GSB\tiny{03} & 1.3907 & 1.4049 & 0.9846 & 0.4588 & 0.1494 & \textbf{0.0355} & 0.0064 & 0.0018 & 0.0007 & 0.0003 & 0.0001 & 0.0001\\

        SHU & 1.6323 & 1.1999 & 0.6185 & 0.221 & \textbf{0.058} & 0.0131 & 0.0032 & 0.0013 & 0.0006 & 0.0003 & 0.0001 & 0.0001\\

        CS & 1.9331 & 0.4205 & 0.284 & \textbf{0.0856} & 0.0256 & 0.0073 & 0.0055 & 0.0024 & 0.0009 & 0.0006 & 0.0004 & 0.0002\\

        Z & 1.3907 & 1.4049 & 0.9846 & 0.4588 & 0.1494 & \textbf{0.0355} & 0.0064 & 0.0018 & 0.0007 & 0.0003 & 0.0001 & 0.0001\\

        SHR & 1.1192 & 1.4395 & 1.3892 & 0.8548 & 0.3485 & \textbf{0.1035} & 0.0229 & 0.0051 & 0.0015 & 0.0006 & 0.0003 & 0.0001\\

        B & 1.0439 & 1.149 & 0.5166 & 0.0477 & 0.4098 & 0.7663 & 0.6473 & 0.3188 & \textbf{0.1154} & 0.0366 & 0.0115 & 0.0039\\

        CW & 1.9216 & 0.4848 & 0.2332 & \textbf{0.0741} & 0.0225 & 0.0074 & 0.0027 & 0.0011 & 0.0005 & 0.0002 & 0.0001 & 0.0001\\
        
        \textbf{Bayesian} & & & & & & & & & & & & \\ 

        PYM & NaN & Inf & Inf & 0.7068 & 0.1991 & \underline{0.0358} & 0.0513 & 0.121 & 0.2211 & \underline{0.3333} & 0.6226 & 1.0733\\

        BAY & 1.0186 & 1.1612 & 1.0402 & 0.674 & 0.3274 & \textbf{0.1241} & 0.037 & 0.0105 & 0.003 & 0.0009 & 0.0003 & 0.0001\\

        LAP & 1.0832 & 1.3259 & 1.2046 & 0.7027 & 0.2638 & \textbf{0.0613} & 0.0099 & 0.0054 & 0.0041 & 0.0027 & 0.0018 & 0.0011\\

        JEF & 1.0627 & 1.2693 & 1.1387 & 0.6822 & 0.2806 & \textbf{0.0794} & 0.0147 & 0.003 & 0.0011 & 0.0006 & 0.0004 & 0.0002\\

        SG & 1.0372 & 1.1904 & 1.0609 & 0.6804 & 0.3274 & \textbf{0.1233} & 0.0366 & 0.0104 & 0.0029 & 0.0009 & 0.0003 & 0.0001\\

        MIN & 1.0604 & 1.2555 & 1.1377 & 0.7175 & 0.3267 & \textbf{0.113} & 0.0293 & 0.0068 & 0.0017 & 0.0007 & 0.0005 & 0.0004\\

        NSB & 1.0818 & 1.3518 & 1.2607 & 0.7297 & 0.2473 & \textbf{0.048} & 0.0057 & 0.004 & 0.0028 & 0.0011 & 0.0004 & 0.0001\\

        ANSB & NaN & NaN & NaN & 5.5632 & 9.7871 & 15.6916 & 22.932 & 31.1686 & 40.1334 & 49.8185 & 60.2975 & 71.6383\\ 
        \hline
    \end{tabular}
    \end{adjustbox}
    \caption{Mean Squared Error (MSE) of mutual information (MI) estimations (domain size $k=256$). The values in bold inform about the sample size from which the MSE, for a given estimator, starts to stabilise under a threshold of 0.1.}
    \label{tab:mse_delta_mi_256}
\end{table*}

It follows that ``convergence stability'' is a valuable quality when selecting an estimator, especially the point where this stability is achieved. We summarise the advantages of using an estimator that provides stable estimations with the least amount of data as follows:

\begin{itemize}
    \item [-] \textit{Efficiency}: Given a relatively small sample, the chance of achieving the same performance (within a margin of accepted fluctuation) that you would get with larger samples is higher compared to using other estimators. Ideally, the size of this sample should coincide with the flattening-off point. This conclusion is based on the strong assumption that the estimation error tends to reduce for larger samples.
    \item [-] \textit{Scalability}: The data collection effort is minimised, ensuring that even with larger samples, the quality of the estimation does not improve.
    \item [-] \textit{Robustness}: The estimator is robust against fluctuations (variability) in the estimation error caused by changes in the observed data structure.
\end{itemize}

Our analysis is conducted for the three Shannon quantities of interest, in the context of different domain sizes ($k$). For the sake of simplicity, details on the delta analysis are shown only for the MI scenario with $k=$ 256 (see~\Cref{tab:mse_delta_mi_256}). The safe sample sizes (\emph{Fp}) marking the beginning of the plateau area of the MSEs curve are shown, for each estimator, in~\Cref{tab:Fp_tab}.\hide{respectively in case of small and high ground truth. We will call the latter \emph{flattening-off point curve (Fc)}.} \hide{Six scenarios of PD defined on a domain of sizes 256, 1024, 4096, 16384, 65536, 262K have been considered.}

Our analysis is conducted for the three Shannon quantities of interest in the context of different domain sizes ($k$). For simplicity, details on the delta analysis are shown only for the MI scenario with $k = 256$ (see~\Cref{tab:mse_delta_mi_256}). The safe sample sizes (\emph{Fp}) marking the beginning of the plateau area of the MSE curves are shown for each estimator in~\Cref{tab:Fp_tab}.

In the following sections, we delve into the examination of the flattening-off points of the 18 entropy estimators for each Shannon measure, given different domain sizes. Our purpose is to identify the estimators that achieve earlier stable convergence. \hide{Due to space constraints}For better visual clarity, only a subset of the analysed scenarios will be graphically illustrated.

\textit{Entropy} (H) - As evidenced in~\Cref{tab:Fp_tab_h}, there exists a strong similarity among specific subsets of entropy estimators in terms of convergence speed and stability. Rows having the same colour reflect a perfect match of flattening-off points for the corresponding estimators across different $k$ scenarios. This means their estimations become independent of the sample size under the same conditions. The darker the colour, the faster the overall convergence for a given estimator. Particularly noteworthy are the Chao-Shen (CS) and Chao-Wang-Jost (CW) estimators, which jointly stand out prominently for their superior consistency across varying domain sizes. This is less evident for smaller input spaces. In fact, the $Fp$ for CS and CW (in purple) is smaller than all the other estimators for domains larger than $\approx$4000. 
To pick just some explanatory cases, for a probability distribution defined on a domain of 4096 values, CS and CW estimations exhibit negligible fluctuations ($<0.1$) when analysing samples larger than $2^{10}$ (1024). They are followed by the Grassberger (GSB\textsubscript{88,03}), Schurmann (SHU), Miller-Madow (MM), and Zhang (Z) estimators (in green), which, for the same domain size, stabilise with samples larger than $2^{11}$ (2048). All the remaining estimators show worse performance. As the size of the domain increases, the convergence becomes slightly slower, but CS and CW consistently perform well.
We validate these observations by measuring, for each estimator, the slope and Euclidean distance of the set of flattening-off points for all the values of $k$ analysed, with respect to the optimal case. The ideal scenario is where the estimations show complete independence from the sample size (i.e., flatten out at the smallest sample size in our range $2^3$) for every $k$. This translates into comparing the current $Fp_s$, e.g., $[9,9,10,9,10,12]$ for CW, with $[3,3,3,3,3,3]$.

The results in~\Cref{tab:Fp_tab_h} confirm that CS and CW are dominant, especially for larger $k$ where the other estimators fail to stabilise within the range of sample sizes studied. The lowest slope (0.5) for CS and CW indicates that the rate of variation of the flattening-off points for increasing domain sizes is relatively small. This means that their convergence speed is not significantly affected (slowed down) by the domain size. The lowest Euclidean distance (24.23) shows, in turn, that a stable performance is achieved with fewer samples compared to the other estimators. Both measurements imply that CS and CW can be particularly effective at producing reliable and consistent results even for relatively small datasets, compared to other estimators. Namely, they can extract maximum information from limited data, delivering the same performance you would get with larger samples, yet using smaller samples.
Following them are GSB\textsubscript{88,03}, SHU, MM, and Z with a slope and Euclidean distance of 1.33 and 30.0. These values confirm that their estimations converge slightly faster than CS and CW and require slightly larger samples. On the other hand, notably weak are the B, PYM, NSB, and ANSB estimators, which do not converge within the sample size range studied. The study of the convergence's goodness (accuracy) is deferred to~\Cref{sec:accuracy}.

\begin{figure}[t!]
\centering
    \begin{minipage}[b]{\columnwidth}
    \centering
        \begin{subfigure}[b]{0.45\columnwidth}
        \includegraphics[width=\columnwidth]{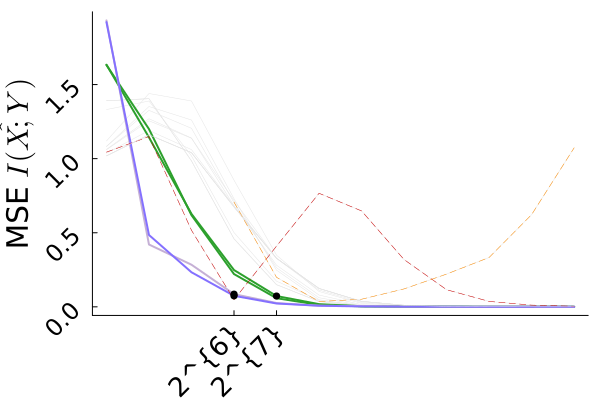}
            \caption{$k=256$}
            \label{fig:right1}
        \end{subfigure}
        \begin{subfigure}[b]{0.45\columnwidth}
        \includegraphics[width=\columnwidth]{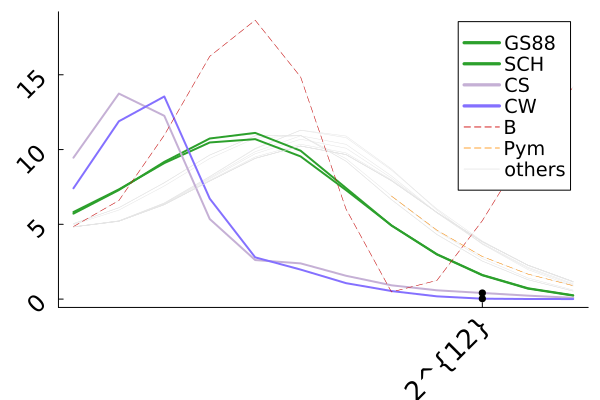}
            \caption{$k=65536$}
            \label{fig:right2}
        \end{subfigure}
    \end{minipage}\\
    \caption{Mean squared error of the estimations for the MI. The fastest converging methods are highlighted with bold colored lines: CS-CW (purple) and GSB\textsubscript{88}-SHU (green). Only two values for $k$ are selected to illustrate the performance in boundary scenarios: $k=256$, column (a) and $k=65536$, column (b). Estimators exhibiting outlier behaviour (B-PYM) are shown with dashed red-orange lines. Both the sets are consistent across all three Shannon metrics, indicating uniform performance patterns. For clear visual interpretation, the ANSB estimator is not displayed due to its exponential errors.}
    \label{fig::MSE-Fp_graph}
\end{figure}

\textit{Mutual Information} (MI)
Estimates of the mutual information exhibit a nearly identical convergence trend to that observed for the entropy (in the previous section), as shown in~\Cref{tab:Fp_tab_mi} and~\Cref{fig::MSE-Fp_graph}. However, there are two minor differences. First, CS and CW estimators outperform in all the varying domain scenarios under analysis, including the smallest ones, contrary to H. Second, CW is marginally faster than CS for $k \geq 16384$. However, it shows slightly slower convergence for larger $k$ relative to H. Specifically, for a probability distribution defined on a domain of 65536 values, its estimations stabilise with samples larger than $2^{12}$, compared to $2^{10}$ in the H scenario. A key observation here is how the two groups of estimators represented by CS-CW and GSB\textsubscript{88}-SHU consistently outperform across different Shannon quantities, albeit with slightly less prominence for GSB\textsubscript{88}-SHU in the H scenario. Both groups graphically dominate (from below) all the other estimators, with a greater emphasis on the first group, where CW stands out for larger $k$. As observed also in the H scenario, CS and CW are dominant especially for larger $k$, where the other estimators fail to stabilise within the range of sample sizes studied. Therefore, overall MI estimations obtained using the Chao-Shen and Chao-Wang-Jost methods are more consistent (stable) against the sample size, across growing domain sizes, compared to the other estimators. In a nutshell, these estimators require fewer samples to achieve their optimal performance. Moreover, similarly to the entropy estimation, the Bonachela (B), PYM, and ANSB methods show overall unstable performance with increasing sample size.

Finally, we interestingly notice that the estimation trends for GSB\textsubscript{88}-SHU estimators are well aligned with all the estimators except for CS-CW (and obviously the under-performing Bonachela, PYM, and ANSB). This alignment is clearly visible in the figure. As $k$ increases, the performance gap between CS-CW and all the other estimators becomes more pronounced. This highlights the superior stability of the CS-CW estimators for larger domains, as observed earlier.

\textit{Conditional Mutual Information} (CMI) - Focusing on the conditional mutual information, the results of our analysis (see~\Cref{tab:Fp_tab_cmi}) confirm the conclusions derived from the previous observations about H and MI estimations: CS-CW are the fastest converging estimators, followed by GSB\textsubscript{88} and SHU, while B, PYM, and ANSB are the most unstable and under-performing ones. Therefore, the overall MSE trend for the 18 estimators is consistent with that observed in the context of the other two information-theoretic quantities, with greater proximity to the MI scenario. However, unlike the latter, other estimators closely approach the performance of the optimal CS-CW methods (even for relatively large $k$). Accordingly, this translates into a reduction in the gap between the performance of CS-CW and that of the other estimators. Nevertheless, the CS and CW estimators continue to exhibit superior performance, with CW showing a slight edge (though less pronounced than in the mutual information (MI) scenario for large $k$).

\noindent
\fbox{\begin{minipage}{\columnwidth}
\textbf{Answer to RQ1:} 

We discover that the Chao-Shen (CS) and Chao-Wang-Jost (CW) estimation methods enable a significant reduction in the data collection effort, without compromising their performance. They prove to be the fastest converging estimators (i.e., require fewer samples to achieve reliable estimates, compared to the other estimators), with CW standing out prominently in large domain scenarios. Notably, these estimators consistently perform well across the three information-theoretic metrics analysed (H, MI, CMI). They are followed by the Grassberger (GSB\textsubscript{88}) and Schurmann (SHU) estimators.
\end{minipage}}

\begin{table}[h!]
\centering
    \begin{adjustbox}{width=0.7\columnwidth}
    \renewcommand{\arraystretch}{1.2} 
    \newcolumntype{L}{>{}^w{c}{6mm}} 
    \begin{tabular}{ >{\bfseries}l | L L L | L L L | L L L L }
        \hline
         k & \multicolumn{3}{c}{\textbf{ H}} & \multicolumn{3}{c}{\textbf{ MI}} & \multicolumn{4}{c}{\textbf{ CMI}} \\
          & \makecell{GSB\tiny{88}\\SHU} & CS & CW & \makecell{GSB\tiny{88}\\SHU} & CS & CW & GSB\tiny{88} & SHU & CS & CW \\
         \hline
         256 & 2.0 & 2.0 & 2.0 & 0.5 & 0.25 & 0.25 & 0.25 & 0.5 & 0.25 & 0.25 \\[0.6em]
         
         1024 & 0.5 & 0.5 & 0.5 & 0.5 & 0.25 & 0.13 & 0.5 & 0.5 & 0.13 & 0.13 \\[0.6em]
          
         4096 & 0.5 & 0.25 & 0.25 & 0.5 & 0.13 & 0.13 & 0.5 & 0.5 & 0.25 & 0.06  \\[0.6em]
          
         16384 & 0.5 & 0.03 & 0.03 & 0.5 & 0.13 & 0.06 & 0.5 & 0.5 & 0.25  & 0.5 \\[0.6em]
          
         65536 &  & 0.02 & 0.02 & & & 0.06 &  &  &  & 0.06 \\[0.6em]
         
         262K &  & 0.02 & 0.02 & &  & 0.03 &  & &  &  \\
    \hline
    \end{tabular}
    \end{adjustbox}
    \caption{Ratios of ``safe'' sample size (\emph{Fp}) to domain size ($k$) for the four fastest converging estimators of entropy $H$, mutual information $I(X;Y)$, and conditional mutual information $I(X;Y|Z)$. The values show the sample size at which the estimations stabilise as a proportion of the domain size.}
    \label{tab:ratios}
\end{table}

\subsection{Sample Size}

In the previous section, we identified four estimators showing early stable convergence (CS-CW, GSB\textsubscript{88}-SHU) by determining the corresponding flattening-off points ($Fp$). We remind that an $Fp$ denotes the minimal amount of data needed to obtain consistent estimations, meaning that further increasing the sample size yields negligible improvements in accuracy. In this section, we delve deeper into their analysis by offering a more rigorous interpretation. Essentially, we express the ``safe'' sample sizes ($Fp_s$) of each estimator as a proportion of the domain size ($k$). This aims to provide more practical guidelines for practitioners, helping them understand how to minimise data collection effort without significantly compromising the performance of a given estimator, based on specific context and requirements. Compared to absolute values, ratios offer more informative and standardised context-sensitive measures. By using ratios, it becomes easier to interpret how the sample size requirements scale with the variability in the data (domain size). Furthermore, ratios enable more effective comparisons across different scenarios and domain sizes, thereby facilitating better decision-making and resource planning in practical applications.

Our findings (see~\Cref{tab:ratios}) reveal consistent patterns in the ratios of ``safe'' sample size to domain size ($Fp/k$) observed across different values of $k$. This is particularly evident for MI estimations. The proportion remains constant at 0.5 for the GSB and SHU estimators, indicating that the safe sample size can always be confidently considered as half of the domain size. This consistency suggests a linear relationship between the sample size and the domain size. 

For CS and CW, the ratio decreases for larger $k$. This drop is sharper for CW, confirming its pronounced efficiency with larger domains. Focusing on the CMI, the linearity characterising the GSB and SHU ratios is less strong compared to that in the MI scenario, though still evident. By looking at the table, we notice that it now also extends to CS, whereas we still observe decreasing ratios for CW. However, the exponential relationship between \emph{Fp} and $k$ shows a slightly more unstable decay than in MI.

\noindent
\fbox{\begin{minipage}{\columnwidth}
\textbf{Answer to RQ2:} 
Our analysis reveals that the Chao-Shen (CS) and Chao-Wang-Jost (CW) methods exhibit an exponential decay relationship between the ``safe'' sample size and the domain size, in contrast to Grassberger (1988) and Schumann, which are characterised by a linear relationship between these two variables. Consequently, the amount of data required by CS and CW to achieve their best estimation decreases, on average, exponentially with increasing $k$. The results confirm that the CS and CW estimators are preferable as they perform well with less data, with CW being notably superior.
\end{minipage}}

\begin{table*}[t!]
\footnotesize
\begin{tabular}{l l}
    \begin{subtable}{0.55\textwidth}
    \begin{adjustbox}{width=\textwidth}
    \begin{tabular}{ >{\bfseries}l@{\hskip 0.05in} | c@{\hskip 0.1in} c@{\hskip 0.1in} c@{\hskip 0.1in} c@{\hskip 0.1in} | c@{\hskip 0.1in} c@{\hskip 0.1in} | c@{\hskip 0.1in} c@{\hskip 0.1in} c@{\hskip 0.1in} c@{\hskip 0.1in} }
        \hline
         \multirow{2}{*}{k} & \multicolumn{4}{c}{ \multirow{2}{*}{\textbf{H}}} & \multicolumn{2}{c}{ \multirow{2}{*}{\textbf{MI}}} & \multicolumn{4}{c}{ \multirow{2}{*}{\textbf{CMI}}} \\
         & & & \\
         \hline
         \multirow{2}{*}{256} & MM & LaP & SG & J & \cellcolor{gray!30}CS & \cellcolor{gray!30}CW & MM & ML & Bay & SG \\
          & 12.58 & 9.57 & 9.65 & 12.04 & \color{gray}1.8 & \color{gray}1.79 & 1.25 & 1.24 & 1.24 & 1.24 \\[0.6em]
          
         \multirow{2}{*}{1024} & MM & LaP & SG & J & \cellcolor{gray!30}CS & \cellcolor{gray!30}CW & \cellcolor{gray!15}GS\textsubscript{88} & \cellcolor{gray!15}SHU & &  \\
          & 22.21 & 20.82 & 20.98 & 23.64 & \color{gray}5.06 & \color{gray}5.18 & \color{gray}3.04 & \color{gray}3.16 \\[0.6em]
          
         \multirow{2}{*}{4096} & MM & & \cellcolor{gray!15}GS\textsubscript{88} & \cellcolor{gray!15}SHU & \cellcolor{gray!30}CS & \cellcolor{gray!30}CW & \cellcolor{gray!15}GS\textsubscript{88} & \cellcolor{gray!15}SHU & & \cellcolor{gray!30}CW  \\
          & 38.86 & & \color{gray}38.89 & \color{gray}38.87 & \color{gray}12.75 & \color{gray}12.63 & \color{gray}6.94 & \color{gray}7.23 &  & \color{gray}6.85 \\[0.6em]
          
         \multirow{2}{*}{16384} & \cellcolor{gray!30}CS & \cellcolor{gray!30}CW & \cellcolor{gray!15}GS\textsubscript{88} & \cellcolor{gray!15}SHU & \cellcolor{gray!30}CS & \cellcolor{gray!30}CW & \cellcolor{gray!15}GS\textsubscript{88} & \cellcolor{gray!15}SHU & \cellcolor{gray!30}CS  & \cellcolor{gray!30}CW \\
          & \color{gray}60.91 & \color{gray}56.36 & \color{gray}61.8 & \color{gray}61.84 & \color{gray}25.58 & \color{gray}24.68 & \color{gray}13.29 & \color{gray}13.8 & \color{gray}13.55 & \color{gray}13.33 \\[0.6em]
          
         \multirow{2}{*}{65536} & \cellcolor{gray!30}CS & \cellcolor{gray!30}CW & \cellcolor{gray!15}GS\textsubscript{88} & \cellcolor{gray!15}SHU & \cellcolor{gray!30}CS & \cellcolor{gray!30}CW & \cellcolor{gray!15}GS\textsubscript{88} & \cellcolor{gray!15}SHU & \cellcolor{gray!30}CS & \cellcolor{gray!30}CW \\
          & \color{gray}80.48 & \color{gray}77.27 & \color{gray}97.66 & \color{gray}97.67 & \color{gray}44.83 & \color{gray}42.45 & \color{gray}23.79 & \color{gray}24.56 & \color{gray}21.64 & \color{gray}18.83 \\[0.6em]
          
         \multirow{2}{*}{262K} & \cellcolor{gray!30}CS & \cellcolor{gray!30}CW &  &  & \cellcolor{gray!30}CS & \cellcolor{gray!30}CW & \cellcolor{gray!15}GS\textsubscript{88} & \cellcolor{gray!15}SHU & \cellcolor{gray!30}CS & \cellcolor{gray!30}CW \\
          & \color{gray}105.2 & \color{gray}103.6 &  & & \color{gray}72.14 & \color{gray}67.3 & \color{gray}38.1 & \color{gray}39.18 & \color{gray}34.48 & \color{gray}30.87 \\
    \hline
    \end{tabular}
    \end{adjustbox}
    \caption{}
    \label{tab:acc_tab_allN}
    \end{subtable} &
    \begin{subtable}{0.418\textwidth}
    \begin{adjustbox}{width=\textwidth}
        \begin{tabular}{>{\bfseries}l@{\hskip 0.05in} | c@{\hskip 0.05in} c@{\hskip 0.05in} | c@{\hskip 0.05in} c | c@{\hskip 0.05in} c@{\hskip 0.05in} | c@{\hskip 0.1in} c@{\hskip 0.05in} }
        \hline
          \multirow{2}{*}{k} & \multicolumn{4}{c}{\textbf{H}} & \multicolumn{4}{c}{\textbf{CMI}} \\
           & \multicolumn{2}{c}{$<Fp$} & \multicolumn{2}{c}{$>=Fp$} & \multicolumn{2}{c}{$<Fp$} & \multicolumn{2}{c}{$>=Fp$} \\
         \hline
         \multirow{2}{*}{256} & LaP & SG &  SHU & GS88 & \color{gray}ML & \color{gray}Bay & \cellcolor{gray!30} CW & \cellcolor{gray!30}CS\\
          & J & MM & & & \color{gray}SG & \color{gray}Min \\[0.6em]
          
         \multirow{2}{*}{1024}  & LaP & SG &  \cellcolor{gray!30}CW & SHU & \color{gray}MM & \color{gray}Z & \cellcolor{gray!30}CW  & \cellcolor{gray!30}CS\\
         & MM & J & & & \color{gray}GS03 & \color{gray}ML \\[0.6em]
          
         \multirow{2}{*}{4096} & MM & \color{gray}GS88 & \cellcolor{gray!30}CW & \cellcolor{gray!30}CS & GS88 & \color{gray}MM & \cellcolor{gray!30}CW & \cellcolor{gray!30}CS \\
         & \color{gray}SHU & LaP & & & \color{gray}GS03 & \color{gray}Z  \\[0.6em]
          
         \multirow{2}{*}{16384}  &  \color{gray}BN & CW & \cellcolor{gray!30}CW & \cellcolor{gray!30}CS & CW & GS88 & \cellcolor{gray!30}CS & SHU \\
          & \color{gray}GS88 & \color{gray}SHU & & & \color{gray}SHU & CS  \\[0.6em]
          
         \multirow{2}{*}{65536} & CW & CS & \cellcolor{gray!30}CS & \cellcolor{gray!30}CW & CW & GS88 & \cellcolor{gray!30}CW & \cellcolor{gray!30}CS \\
          & & &  & & CS & \color{gray}ML \\[0.6em]
          
         \multirow{2}{*}{262K} & CW & CS & \cellcolor{gray!30}CS & \cellcolor{gray!30}CW & & & &\\
          & & &  & & & & \\
    \hline
    \end{tabular}
    \end{adjustbox}
    \caption{}
    \label{tab:acc_tab_Fp}
    \end{subtable}
    \end{tabular}
    \caption{a) Estimation accuracy for entropy (H), mutual information (MI) and conditional mutual information (CMI), illustrated for different domain sizes ($k$), averaged across all samples sizes in range $[2^3-2^{14}]$. The smaller the value, the higher the accuracy. b) Averaged estimation accuracy for entropy (H) and conditional mutual information (CMI) over samples smaller and bigger than the flattening-off points (\emph{Fp}) of the fastest converging estimation methods (CS-CW).}
    \label{tab:acc_tab}
\end{table*}

\subsection{Accuracy}\label{sec:accuracy}

The study of the estimations' convergence speed does not inherently offer strong insights into their level of proximity to the ground truth. An earlier convergence does not imply that the estimations are the most accurate, even for samples smaller than the safe sample size ($Fp$). To complement the findings illustrated in the previous sections, it is imperative to introduce a measure of quality (effectiveness). Therefore, we evaluated the averaged accuracy of the estimators by calculating the area under the curve (AUC) for the estimations observed over increasingly larger sample sizes.

The results (illustrated in~\Cref{tab:acc_tab_allN}) clearly show the superior performance (in terms of accuracy) of the Chao-Shen and Chao-Wang-Jost estimators, particularly in the mutual information (MI) scenario. However, this is less evident for smaller domains when considering both H and CMI estimations. In such cases, other estimators, particularly Miller-Madow, Grassberger (1988), and Schurmann, either closely approach or are more accurate than CS and CW for $k \leq 4096$. These results confirm the conclusions made in the previous sections that CS and CW are notably effective in large domain scenarios.

We interpret this inconsistency as a direct consequence of the presence of unstable peaks in the estimations, mostly for small sample sizes (as seen in~\Cref{fig::MSE-Fp_graph}). This can result in deviations in the estimate accuracy when averaged across samples of different sizes. Standard averaging processes fail to adequately smooth and mitigate these distortions. Consequently, we decided to focus the analysis only on larger samples. Specifically, we measured the average accuracy for samples greater than the ``safe'' sample size identified for CS and CW (see~\Cref{tab:acc_tab_Fp}). This approach dramatically shifts the results, reaffirming the prominence of CS and CW in these scenarios. This data provides strong evidence that CS and CW maintain their effectiveness across the different information-theoretic quantities for relatively larger samples. For samples smaller than the point where their estimations stabilise, CS and CW are more accurate than the other estimators in the context of relatively large input spaces.

\noindent
\fbox{\begin{minipage}{\columnwidth}
\textbf{Answer to RQ3:} 
Our analysis shows that the most accurate estimators tend to be those that also converge quickly. Chao-Shen (CS) and Chao-Wang-Jost (CW) consistently provide more accurate estimations for entropy (H), mutual information (MI), and conditional mutual information (CMI) compared to the other estimators. This occurs across all domain sizes under analysis for relatively larger sample sizes, and across the sample sizes for larger domains.
\end{minipage}}

\hide{CS and CW consistently exhibit superior accuracy for smaller domain sizes and lower ground truth (GT) values, with CS outperforming CW. However, for larger domain sizes, the trend reverses for smaller samples, with Grassberger and Schurmann estimators appearing to become more favorable in such cases. The same trend is observed also for higher GT values, but with CW outperforming. The only notable deviation occurs in the case of CMI, lower GT, and smaller domain sizes, where the performance of CS and CW appears to decline earlier, compared to the MI scenario, when considering smaller samples.}

\section{Discussion}

Entropy estimation is a fundamental aspect of information theory, with significant applications in various fields, including software engineering. Accurate estimation of entropy and related measures such as mutual information (MI) and conditional mutual information (CMI) is crucial for tasks like feature selection, information leakage detection, and data compression. This study compares the performance of 18 widely employed entropy estimators. We evaluate their accuracy and convergence speed across different domain sizes and sample sizes, providing concrete recommendations for software engineers.

\subsection{Comparison of the Different Estimators}

In terms of \textbf{accuracy}, Chao-Shen (CS) and Chao-Wang-Jost (CW) estimators demonstrate superior accuracy across a range of domain sizes and sample sizes. Their mean squared error (MSE) is consistently lower compared to other estimators, particularly for larger domains. This makes them highly reliable for estimating entropy, MI, and CMI in scenarios where the underlying data distribution is unknown. For large domains, both CS and CW maintain high accuracy even with limited data. For
smaller domains, while other estimators like Miller-Madow (MM), Grassberger (GSB88), and Schurmann (SHU) can match or exceed their accuracy, CS and CW still perform robustly. 

The accuracy for the other estimator varies. Miller-Madow (MM) exhibits good accuracy for smaller domains but tends to underperform as the domain size increases. Grassberger (GSB88) and Schurmann (SHU) are effective for smaller domains but are surpassed by CS and CW for larger domains. Finally, Bonachela (B), PYM, and ANSB show instability with increasing sample size and domain size, resulting in lower accuracy.

\begin{figure}[t]
    \centering
    \includegraphics[width=0.45\linewidth]{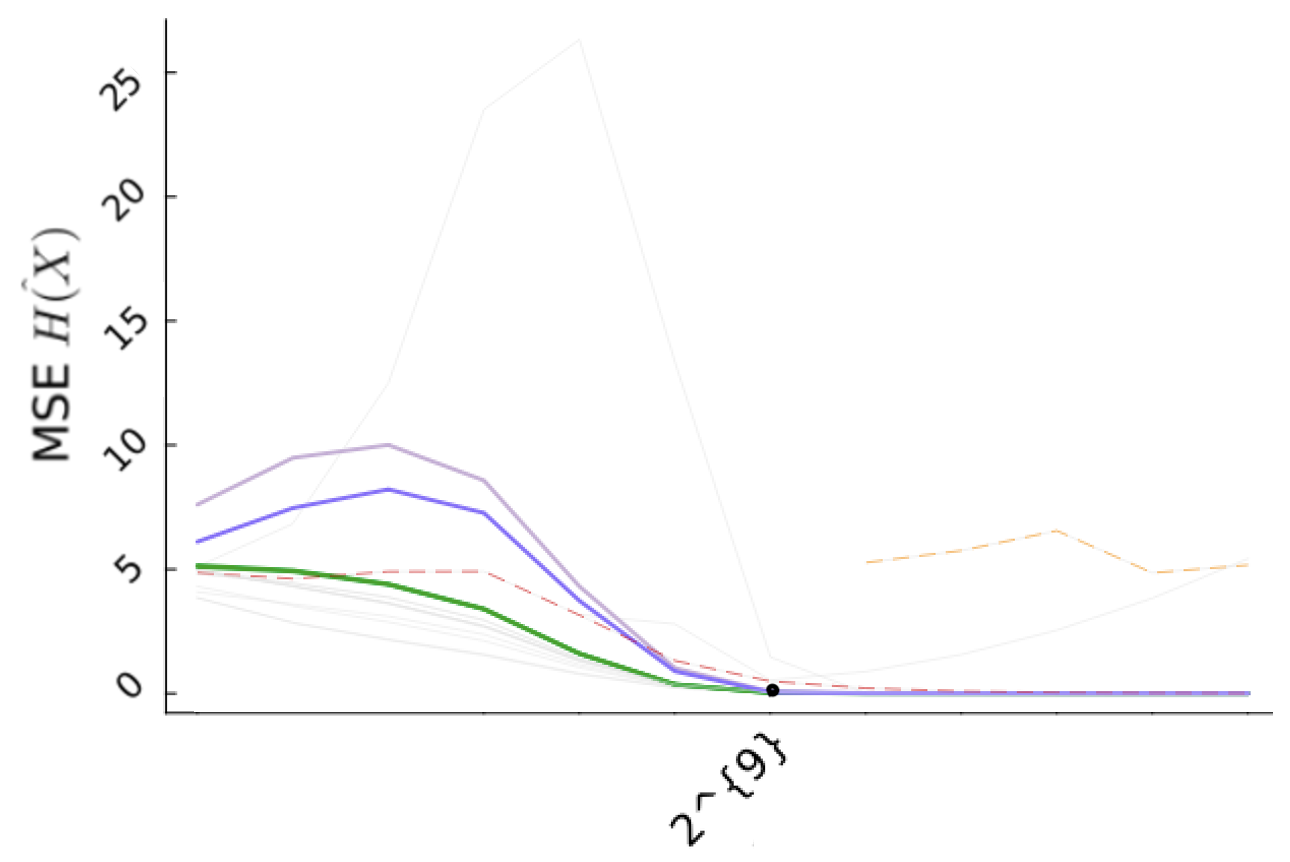}
    \includegraphics[width=0.45\linewidth]{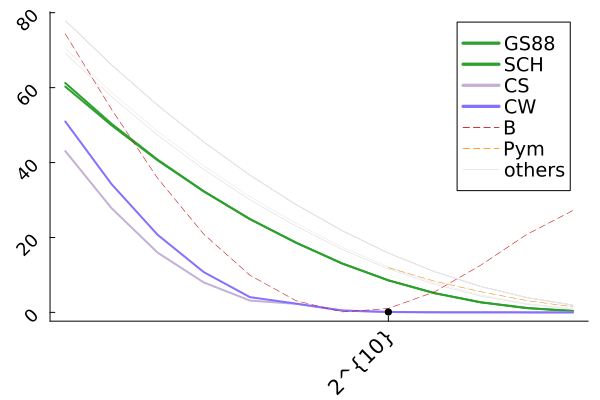}
    \includegraphics[width=0.45\linewidth]{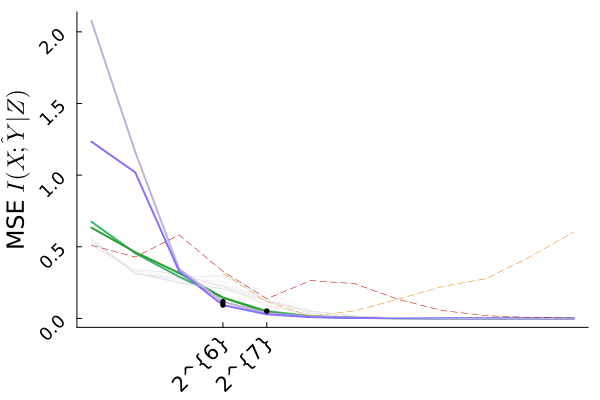}
    \includegraphics[width=0.45\linewidth]{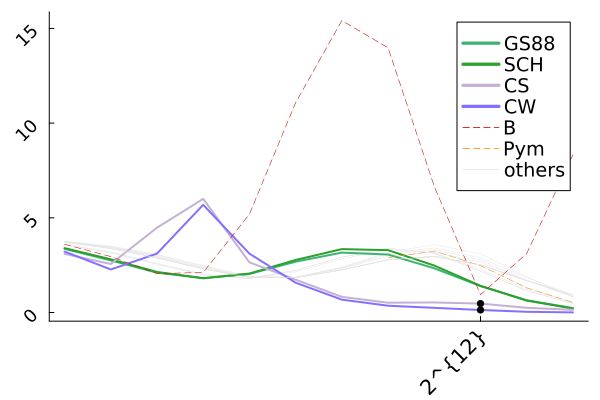}
    
    \caption{Additional plots for entropy (row 1) and CMI estimations (row 2), with $k=256$ (column 1) and $k=65536$ (column 2).}
    \label{fig:addplots}
\end{figure}

From \autoref{tab:mse_delta_mi_256}, for small samples ($\leq 2^8$), CS and CW show the smallest mean squared error compared to the other estimators (less evident for CMI tasks). This error gradually decreases (indicating higher accuracy) for larger samples, with a significantly noticeable gap compared to the other estimators. Bonachela (B) appears to perform better for $n=2^6$, but this is an isolated case due to its tendency to fluctuate; in fact, its MSE drastically increases for $n=2^7$. For large samples ($\geq 2^9$), CW proves to be more robust, although the gap with the other estimators becomes increasingly less evident as the sample size increases (more noticeable in smaller domains). Initially, GSB88 and SHU approach first, followed by MM, GSB03, and Z, and then the rest for larger samples (see \autoref{fig:addplots}).

With respect to the \textbf{convergence speed}, Chao-Shen (CS) and Chao-Wang-Jost (CW) estimators are notable for their rapid convergence, requiring fewer samples to achieve reliable estimates (see \autoref{tab:Fp_tab_cmi}). The ratio of ``safe'' sample size, or flattening-off point (\emph{Fp}), to domain size ($k$) for these estimators decreases exponentially with increasing domain size, highlighting their efficiency (see \autoref{tab:ratios}). Also, CW converges slightly faster than CS in very large domain scenarios (e.g., domains of size 4,096 and above), in line with the observations on its accuracy. 

The convergence speed for the other estimators also differs. Grassberger (GSB88) and Schurmann (SHU) estimators also converge relatively quickly but at a slower pace than CS and CW, especially for large domains. Miller-Madow (MM), GSB03 and Z converge at a moderate rate, especially for MI and CMI tasks, but with less stability than CS and CW. Finally, Bonachela (B), PYM, and ANSB often fail to converge within the sample size range studied, making them less reliable for practical applications.

\subsection{Recommendations for Applying Entropy Estimators}

Based on our findings, we have collected different specific recommendations for software engineers.

\textbf{Selection of Estimators:} Chao-Shen (CS) and Chao-Wang-Jost (CW) estimators are recommended for their consistency and accuracy across various domain sizes and information measures (entropy, mutual information, and conditional mutual information). They tend to converge quickly to the ground truth and maintain effectiveness across larger sample sizes, which is crucial when data collection is resource-intensive. Grassberger (GSB88, GSB03), Schurmann (SHU), Miller-Madow (MM), and Zhang (Z) are also reliable, particularly for medium to large domain sizes. They are slightly less efficient than CS and CW but still provide good performance for larger datasets.

\textbf{Convergence Speed:}  Identify the ``safe'' sample size for your chosen estimator, which is the point where the estimation becomes stable and additional data has negligible impact on accuracy. For CS and CW, this is particularly advantageous as they require fewer samples to achieve stable and reliable estimates.

\textbf{Accuracy:} Ensure that the estimator you choose converges quickly to provide reliable estimates with minimal data. The CS and CW estimators consistently show early stable convergence. CS and CW provide more accurate estimations for entropy, MI, and CMI compared to other estimators. Our findings on convergence and accuracy can help software engineers minimize the effort needed for data collection. Since CS and CW require fewer samples to achieve reliable estimates, they are particularly beneficial in scenarios where data collection is resource-intensive.

\textbf{Implementations:} Regularly update your knowledge on the performance of various estimators. The effectiveness of an estimator can vary based on specific application contexts and the characteristics of the data being analyzed. Leverage available implementations and libraries that provide these estimators. We provide the Julia DiscreteEntropy package \cite{kelly2024discreteentropy}, which ensures uniform precision across different estimators, facilitating reliable comparisons and implementations.

\section{Related Work}\label{sec:relwork}
Other studies have examined entropy estimators in tightly controlled settings. 
C. Rodriguez \emph{et al.}~\cite{Rodriguez2021} empirically determine the best entropy estimator for short byte strings with maximum entropy, focusing on uniform distributions. De Gregorio \emph{et al.} examine Markovian sequences~\cite{DeGregorio2024}, while Pinchas~\emph{et al.} study the performance of well-known entropy estimators across three specific classes of distribution in a large-alphabet regime~\cite{Pinchas2024}. No large work known to us investigates estimating other Shannon measures. De Matos \emph{et al.}~\cite{DeMatos2011} conduct a small study on mutual information, involving only 4 estimation methods and considering three different discretization approaches. 

Our research pioneers a holistic evaluation of entropy estimators for different information quantities, beyond just entropy, more useful to the software engineers (mutual information and conditional mutual information). Unlike the literature's focus on parameterized distributions, we develop joint probability distributions that directly deal with these problems and reflect the natural distributions engineers typically face. We thus offer decisive guidance on the most reliable estimators for engineers needing to apply entropy-based measures without prior data distribution knowledge. Additionally, we are consistent in the precision level across diverse estimators' measurements, setting us apart from other studies that vary in floating point precision limits across different languages.

\subsection{Information theory applications to software engineering}

We highlight the most impactful studies that have successfully integrated information-theoretic measures in software engineering, systematically organized by their application area.

\textit{Feature Selection} - Quantifying the dependence between random variables is crucial in machine learning, particularly in feature selection, where irrelevant features are discarded to enhance model accuracy and reduce overfitting. Within this framework, the minimization of mutual information assumes paramount importance~\cite{Chen2018, Beraha2019}. \textit{Fault Localisation} - Zhang \etal~\cite{Zhang2022} use mutual information (MI) to identify root causes of neural network (NN) crashes by measuring the ``crash relevance'' of program entities. It filters out entities with low MI, which are irrelevant to the crash, to shorten long execution traces and facilitate NN input feeding. The \emph{Inference} fault localization algorithm~\cite{Feyzi2019}, applies conditional mutual information (CMI) to implement a feature selection method to isolate suspicious code. CMI quantifies the interdependence of program statements and their mutual effect on the program termination state. Another significant contribution of entropy measurement is in assisting effective fault localization prioritization~\cite{Yoo2013}. \textit{Trustworthy AI} - Monjezi \etal~\cite{Monjezi2023} use the notion of quantitative individual discrimination (QID) based on Shannon and min-entropy metrics to measure fairness defects in deep neural networks, identifying biased layers and neurons. Moreover, MI serves as a foundational concept for the implementation of discriminators that assist in fair and robust model training~\cite{roh2020}. 

\textit{Cryptography and Data Compression} -  Entropy measures are crucial in data compression, as highlighted by Auli-Llinas~\cite{Auli2023}. When entropy is significantly lower than the maximum information content ($log(k)$), it indicates high redundancy, meaning fewer bits are needed for representation compared to fixed-length encoding. Higher source entropy allows for greater compression. Similarly, higher entropy in encryption keys increases their unpredictability and security ~\cite{Gong1990, Vassilev2014, Zolfaghari2022}. \textit{Information Cartography}~\cite{kelly2022software} estimates Thiel's Uncertainty coefficient, using the Chao-Shen estimator, to map information flow inside programs. \textit{Information Leakage} - In side-channel attacks, conditional entropy measures the information an attacker gains about confidential data from the system’s output. Lower remaining entropy indicates less confidential information leaked to the attacker~\cite{Heusser2010, Mesecan2023}. \textit{Malware Detection} - Entropy has proven effective also in malware detection, with Li \etal~\cite{Li2024} using Shannon entropy to assess prediction uncertainty in Android malware detection models, thereby enhancing their accuracy. Higher entropy signifies greater uncertainty and helps differentiate between correct and incorrect predictions.

Within this wide range of entropy applications, noteworthy is its role in predicting the likelihood of failed error propagation \cite{Androutsopoulos2014} and failed disruption propagation \cite{Petke2021}, in measuring software properties (coupling and cohesion) \cite{Allen1999}, as well as in understanding and improving software quality, from testability \cite{Patel2022} to test suite effectiveness (diversity) \cite{Shi2016}.

Extensive cross-sectional research has utilized Information Theory metrics in time series causal analysis, notably through Directed Information (DI) related to Granger Causality~\cite{Amblard2012}. This has proven effective in root cause analysis of industrial system anomalies and telecommunication network issues~\cite{Duan2014, Yuan2014}. Applications of DI extend to neuroscience, as explored by Quinn \etal~\cite{Quinn2011} and Kim \etal~\cite{Kim2011}, and other fields such as economics, psychology, and ecology. Notably, B\"ohme~\cite{Böhme2018} demonstrated the use of non-parametric Chao and Jackknife estimators, inspired by entropy estimation, in biodiversity studies for discovering rare species.

\section{Conclusions}

This study comprehensively compares the performance of 18 well-known entropy estimators under varying sampling conditions within the framework of general probability distributions. Our primary contribution is providing concrete recommendations on the best estimators to choose, in the absence of any knowledge on the data distribution, when adopting information-theory approaches in software engineering applications. The key distinction of our work lies in the extension of this analysis beyond single entropy to include also mutual and conditional mutual information. These measures prove to be more practical for software engineers, who often deal with complex systems where understanding and optimizing the interplay between components is crucial. 

We make sure that our results are more trustworthy compared to existing comparative studies and closely align with software engineering needs. To that aim we use implementations all from the one package, guaranteeing always the same amount of precision for every estimator. Moreover, instead of focusing on manipulating uniformity of the distributions, we create ``un-parameterises'' joint distributions that closely mirror natural distributions, aligning more with real-world scenarios that software engineers encounter. 

Our findings highlight that the Chao-Shen and Chao-Wang-Jost estimators consistently converge more quickly to the ground truth, regardless of domain size or information measure. They also outperform other estimators in terms of accuracy as sample sizes increase, enabling significant reductions in data collection efforts without compromising performance. Therefore, in the absence of distribution knowledge, these estimators prove to be the best choice for reliable estimation across different information measures.


\begin{acks}
Ilaria La Torre and David Clark were partially supported by Meta's Flaky project. David Kelly and Hector Menendez were partially funded by UKRI Trustworthy Autonomous Systems Hub (reference EP/V00784X/1) and Trustworthy Autonomous Systems Node in Verifiability (reference EP/V026801/2). David Kelly was also funded by CHAI - EPSRC AI Hub for Causality in Healthcare AI with Real Data (EP/Y028856/1)
The Alan Turing grant G2027 - MuSE is gratefully acknowledged.
\end{acks}



\bibliographystyle{ACM-Reference-Format}
\bibliography{main}

\end{document}